\documentclass[12pt]{article}
\usepackage[hmargin=2.5cm,vmargin=3cm]{geometry}
 \usepackage{graphicx}
 \usepackage{amsmath}
 \usepackage{amssymb}
  \usepackage{enumerate}
\usepackage{cite}
\usepackage{mathabx}
\usepackage{url}
\newcommand{\bey}{\begin{eqnarray}}
\newcommand{\eey}{\end{eqnarray}}
\usepackage{etoolbox}
\apptocmd{\thebibliography}{\setlength{\itemsep}{1pt}}{}{}

\begin{document}

\title{Gravitational effects in macroscopic quantum systems: a first-principles analysis  }
\author { Charis Anastopoulos\footnote{anastop@physics.upatras.gr}, Michalis Lagouvardos\footnote{mikelagou@upatras.gr}, and   Konstantina Savvidou\footnote{ksavvidou@upatras.gr}   \\
 {\small Department of Physics, University of Patras, 26500 Greece} }

\maketitle

\begin{abstract}
We analyze the weak-field limit of   General Relativity with matter and its possible quantisations. This analysis aims towards a predictive quantum theory  to provide a first-principles description of gravitational effects in macroscopic quantum systems. This includes recently proposed experiments on the generation of (Newtonian) gravitational forces from quantum distributions of matter, and phenomena like gravity-induced entanglement, gravitational cat states, gravity-induced Rabi oscillations, and quantum causal orderings of events. Our main results include: (i) The demonstration that these phenomena do not involve true gravitational degrees of freedom. (ii) We show that, unlike full general relativity, weak gravity with matter is a parameterised field theory, i.e., a theory obtained by promoting spacetime coordinates to `dynamical' variables. (iii) Quantisation via gauge-fixing leads to an effective field theory that account for some phenomena, but at the price of gauge dependence that manifests more strongly on spacetime observables. This ambiguity is a manifestation of the problem of time that persists even in weak gravity. (iv) A consistent quantisation of parameterised field theories is essential for a predictive and spacetime covariant theory of weak gravity that describes gravitational effects in macroscopic quantum systems. We also discuss the implication of our results to gravitational decoherence theories, the notion of locality in gravity vis-a-vis quantum information theory, and the intriguing possibility that proposed solutions to the problem of time can be tested in weak-gravity quantum experiments.
\end{abstract}

\section{Introduction}

Quantum theory and General Relativity (GR) are the two main pillars of modern theoretical physics. Both are extremely successful theories in their respective domain. Nevertheless, their structures are fundamentally incompatible. For example, in quantum theory, the concept of measurement appears to be fundamental, while in GR it is not. Time in GR is dynamical, while in quantum theory it is described as an external predetermined parameter. Finding a unifying theory, a theory of quantum gravity, is one of the most important goals of current research.

Despite intense efforts over the past four decades, there is no functional theory of quantum gravity. This is largely due to the lack of experimental data to guide the theory. Quantum gravitational phenomena are estimated to be significant at the Planck scale, which is well outside our current experimental reach, at least directly. The only potential exception are the quantum signatures in gravitational waves produced during inflation.

The identification of the Planck scale for quantum gravity effects originates from the use of the  de Broglie wave-length as the characteristic length-scale of quantum effects. However, this applies only to specific quantum states that are relevant to a specific class of experiments, for example, scattering experiments. Nowadays, we can prepare particles in states that manifest quantum behavior at mesoscopic or even macroscopic scales \cite{MQP1, MQP2, MQP3, MQP4, MQP5, MQP6, MQP7, MQP8, MQP9, MQP10, MQP11, MQP12}. Schr\"odinger cat states are an example of such states, i.e., quantum superpositions of localised states  for particles of mass $M$, with a macroscopic distance $L$ of  their centers. For such states, the effect of gravity becomes stronger as $L$ and $M$ increase \cite{AnHu15}.

The idea of looking for gravitational effects in macroscopic quantum systems originates from Karolyhazy \cite{Karol}, and it has intensified in recent years. The reason is that we are reaching a level of precision that allows for concrete experimental testing. In this paper, we undertake a first-principles analysis of such systems, by considering the quantisation of linearised gravity interacting with matter. This level of description  suffices for describing the systems under consideration, since the gravitational interaction is weak. The relevant effects can be described  in terms of Newtonian gravity. However, a first-principles analysis that relates the gravitational behavior of macroscopic quantum systems to the spacetime description of GR is crucial for understanding how relevant experiments impact our  fundamental understanding of gravity.

Regarding the effect of gravity on macroscopic quantum systems, research has focused on two contrasting, but complementary, directions. On one hand, there is the idea that gravitational effects lead to decoherence of  cat states with different mass density \cite{Karol, Diosi, Penrose, Penrose2}, typically at a time-scale of the order of the gravitational self-energy. This is a premise of the Diosi-Penrose model for gravity-induced decoherence. There are several other models for gravitational decoherence \cite{PowPer, ReJa, Breuer, Blencowe13, AnHu13}, and a large number of proposed experiments, for example, \cite{MQP3, MQP10, gravdec, HSMY }.

In contrast, if cat states of matter are preserved, then they must generate a gravitational field with fluctuations that mirror the quantum fluctuations of matter. This was first mentioned by Feynman  at the 1957 Chapel Hill Conference---see, Chapter 23 of Ref. \cite{ChapelHill}---in the context of a discussion whether gravity should be quantized.
 Ref. \cite{AnHu15}, provided a concrete description of such states, suggesting the name
  `gravitational cat states' was suggested, and investigating experimental consequences, see, also Ref. \cite{DAH}. In Refs. \cite{Bose17, Vedral17}, it was proposed that  the gravitational interaction of a pair of cat states leads to the generation of entanglement, and that this effect is measurable with contemporary technology. Other quantum resources, like non-Gaussianity, may also be generated \cite{nonG}. In Ref. \cite{BCM18}, the visibility of photonic interference is proposed as a criterion for identifying gravitational phenomena in nanomechanical oscillators, while in Ref. \cite{AnHu20}, gravity-induced Rabi oscillations were proposed as a more particularly persistent feature in macroscopic quantum set-ups. Gravitational effects in quantum gases, including Bose-Einstein condensates were proposed in Refs. \cite{PF, HPF, Haine}.
   In Refs.
 \cite{ZCPB, CGBB}, it was proposed that in the same regime, we can analyze superpositions of different causal orders of spacetime events.

 Refs. \cite{Bose17, Vedral17} claimed that the measurement of gravity-induced entanglement would demonstrate the quantum nature of the gravitational field. This claim has generated intense critique and discussions both against and in support of it, see, for example, Refs. \cite{claims1, claims2, claims3, claims4, claims5, claims6, claims7, claims8, claims9, AnHu20}. This controversy is part of the motivation of the present work. We believe that this issue has a unique and unambiguous answer that can be established from a first-principles analysis.

 However, our most important motivation is to understand the  implications of experiments with macroscopic quantum sources  for quantum gravity theories. For experiments in weak gravitational fields, the relevant classical theory is GR with matter, in the weak field approximation. After we express this theory in a Hamiltonian form, we proceed to its quantisation.
 Since our starting point is GR, the classical description carries the symmetry of spacetime diffeomorphisms. In the Hamiltonian framework, the relevant symmetry is the algebra of surface deformations, the {\em Dirac algebra}, which  is expressed in terms of first-class constraints. The Hamiltonian vanishes because of these constraints. The key challenge in quantisation is to obtain unambiguous predictions and a coherent spacetime picture for the quantum phenomena.

We find  that many conceptual problems associated with the quantisation of gravity  are preserved in the linearised approximation. The incompatibility between GR and quantum theory is structural \cite{Ishamstr}, and it does not go away with approximations.
Brute-force solutions like gauge-fixing provide a workable quantised  description of quantum matter interacting with weak gravity, but at the price of complete ambiguity in defining  spacetime observables.

Experiments with gravitating quantum systems may provide significant insights about gravity. To this end,  we must first understand the predictions made by the default theory of such systems, namely the quantisation of the weak gravity limit of GR.
In this paper, we  map out   difficulties in providing such predictions, and we propose what we believe to be the most natural solution.

Our main results  are the following.

 \smallskip

 \noindent 1. In appropriate gauges, the quantisation of linearised gravity interacting with matter leads to a non-relativistic quantum field theory (QFT) with non-local interactions for the Newtonian gravity limit. This QFT description accounts for gravity-induced entanglement experiments, and it allows for the demonstration that no true gravitational degrees of freedom are involved. The latter are decoupled from this level of description. This implies that no experimental proof of quantum gravity is possible with the currently proposed experiments. Any convincing experimental proof would require access, direct or indirect, to the true degrees of freedom.

  \smallskip

\noindent  2. We  constructed  operators for the spacetime metric that, in principle, could  allow us to analyze fundamental issues like light-cone fluctuations or indefinite causal order in macroscopic quantum systems. However, these operators are constructed in a gauge-dependent way. In general, quantization   via gauge fixing leads to an ambiguous spacetime description {\em even in the Newtonian regime}.

  \smallskip

\noindent  3.  The root of the problem above is the conflict between the symmetry of general covariance and  quantum theory.  This incompatibility  is an aspect of the infamous {\em problem of time} in quantum gravity \cite{Isham92, Kuchar92, Anderson17}. We show here (Sec. 4) that this problem persists even in the weak gravity regime. However, this difficulty is also an opportunity for the experimentalist: our analysis implies that proposed solutions to the problem of time can be tested in macroscopic quantum systems.

  \smallskip

\noindent    4. A key result of our analysis is that weak gravity interacting with matter is a Parameterised Field Theory (PFT) \cite{Kuchar88, Kuchar89, ToVa1, ToVa2}, i.e., a field theory obtained by promoting spacetime coordinates into dynamical fields. The quantisation of PFTs is difficult, and it  leads to quantum theories that strongly differ from usual QFTs. However, if this quantisation can be implemented rigorously, we will obtain a quantum theory that incorporates evolution with respect to all possible foliations. This theory would define {\em unambiguous} spacetime observables for weak gravity, and it would not be equivalent to the effective field theory defined by gauge fixing. Therefore, it would lead to novel testable predictions.

   \smallskip

\noindent   5. An important conclusion is that theories with first-class constraints like GR likely require a reappraisal of quantum informational notions like that of local operations. GR is fully local and causal even if the Hamiltonian of the true degrees of freedom involves non-local terms. The existence of {\em instantaneous laws} \cite{CQG} expressed by the constraints should not be mistaken for non-local behavior. We give an example why a failure to appreciate this point could lead to erroneous assumptions in experiments for gravity-induced entanglement

\smallskip

\noindent   6. The construction of a quantum theory for weakly gravitating matter that (i) is fully predictive and (ii) has an unambiguous spacetime description is challenging. If this challenge is not eventually met, then perhaps we would have to accept Penrose's point about an irreconcilable ambiguity in combining standard quantum theory and GR \cite{Penrose} even at the weak gravity limit, resulting to gravitational decoherence. Our analysis shows that the gravitational self-energy is not responsible for such ambiguities. Hence, if gravitational decoherence exists, its characteristic time-scale is likely different from what the Diosi-Penrose model predicts.

  \smallskip

  The structure of this paper is the following. In Sec. 2, we analyse the 3+1 description of gravity interacting with matter with the usual (naive) linearisation. We identify the relevant variables, their physical interpretation and the reduced state space. In Sec. 3, we implement an improved linearisation procedure that does less violence in the spacetime properties of the system and we prove that linearised gravity with matter is a parameterised field theory. In Sec. 4, we present a quantisation of weak gravity with gauge fixing, we show that the theory in the ADM gauge reduces to the expected effective field theory, and we show that  spacetime observables can be defined but they are gauge-dependent. In Sec. 5, we argue that the best way to quantise weak gravity is by treating it as a parameterised field theory, we explain the general features of such theories, and why they are expected to provide a proper spacetime description. In Sec. 6, we discuss our results and explore their implications about major issues in the field: whether experiments in macroscopic quantum systems can prove the quantisation of gravity, challenges in using quantum information concepts in relation to locality, the relation to theories of gravitational decoherence and the relevance to the detection of gravitational waves and even gravitons.

\section{Classical description of matter weakly interacting with gravity}
In this section, we describe the classical theory for the interaction of linearised gravity with matter.
\subsection{Lagrangian formulation}
The starting point of our analysis is the Lagrangian description of gravity interacting with matter fields. We consider a Lorentzian metric $g_{\mu \nu}$ on a spacetime manifold $M =  \Sigma \times {\pmb R}$. The action is
\begin{eqnarray}
S = \frac{1}{\kappa} \int d^4x \sqrt{-g}R + \int d^4x {\cal L}_{mat}(q_a, \partial q_a),
\end{eqnarray}
 where $R$ stands for the Ricci  scalar associated to the metric, $g = \det g_{\mu \nu}$, and $q_a$  are matter fields. The constant $\kappa = 16 \pi G$, where $G$ is Newton's constant.

The linearisation proceeds by expanding the metric around the Minkowski metric $\eta_{\mu \nu}$,
\begin{eqnarray}
 g_{\mu \nu} = \eta_{\mu \nu} + \lambda \gamma_{\mu \nu}, \label{lagrangianlin}
 \end{eqnarray}
where $\lambda$ is a formal perturbation parameter.

Keeping a finite number of terms in this expansion breaks the diffeomorphism invariance of the original action, resulting to a theory with different symmetries from the original. The details of this calculation is standard, and we omit them here.

We prefer to implement the linearisation in the Hamiltonian formulation of GR.The reason is that this formulation allows for more flexibility. We can implement linearisation in a way that leads to the same results with the procedure of first linearising at the Lagrangian level and then making a Legendre transformation. But also, we can implement linearisation in a way that preserves the fundamental structure of GR, namely, that its Hamiltonian vanishes due to first class constraints; this is impossible if we start from the linearisation at the Lagrangian level.

\subsection{The Hamiltonian  formulation}
In the Hamiltonian formulation,
the phase space $\Gamma$ for matter and gravity consists of points $(h_{ij}, \pi^{ij}, q_a, p^a)$, where $h_{ij}$ is a three metric on a manifold $\Sigma$, $\pi^{ij}$ is the conjugate momentum, $q_a$ are matter fields and $p^a$ their conjugate momenta. The spacetime metric is constructed from $h_{ij}$ as
\begin{eqnarray}
ds^2 = (- N^2 + N_iN^i) dt^2 + 2 N_i dx^i dt + h_{ij}dx^i dx^j, \label{3+1dec}
\end{eqnarray}
where the coordinates $t$ and $x^i$ are selected by a background spacelike foliation, $N$ is the lapse function and $N^i$ is the shift vector.

The symplectic form on the phase space $\Gamma$ is
\begin{eqnarray}
\Omega = \int d^3 x \delta \pi^{ij}(x) \wedge \delta h_{ij}(x) + \delta p^a(x) \wedge \delta q_a(x),
\end{eqnarray}
and the fundamental Poisson brackets are
\bey
\{ h_{ij}({\pmb x}), \pi^{kl}({\pmb x}')\} = (\delta_i^k \delta_j^l + \delta_i^l\delta_j^k) \delta^3({\pmb x}, {\pmb x}'), \hspace{0.5cm} \{q_a({\pmb x}), p^b({\pmb x}')\} = \delta_a^b \delta^3({\pmb x}, {\pmb x}').
\eey

Assuming that the coupling of matter to gravity involves no derivatives of the metric,
the Hamiltonian is
\begin{eqnarray}
H = \int d^3 x [ N(x) {\cal H}(x) + N^i(x) {\cal H}_i(x)], \label{Ham31}
\end{eqnarray}
where ${\cal H}$ is the Hamiltonian constraint and ${\cal H}_i$ is the momentum constraint, defined by
\begin{eqnarray}
{\cal H} &=& \kappa \frac{\pi^{ij}\pi_{ij} - \frac{1}{2} \pi^2}{\sqrt{h}} - \frac{\sqrt{h}}{\kappa} \; {}^3R + {\cal V} (h, q, p), \label{superH}\\
{\cal H}_i &=& -2 \nabla_j \pi^j{}_i + {\cal V}_i(h, q, p). \label{superM}
\end{eqnarray}
${\cal V}$ and ${\cal V}_i$ are the matter contribution to the constraints; they are  local functionals of the  matter field variables and   of  the three-metric. The lapse function $N$ and  the shift vector $N^i$ play the role of Lagrange multipliers in the dynamics; $\kappa = 16 \pi G$.

Not all matter fields have non-derivative couplings with the metric. For example,  the Dirac field couples with derivatives \cite{Dirac, Kibble, Henneaux}. However, the differences from such  couplings originate from the relativistic spin tensor of the matter fields, and amount to the coupling of spin with the Riemann tensor. In the regime of interest, their contribution is expected to be negligible.  Still, an analysis that includes such effects could be interesting in related set-ups, for examples, quantum systems interacting with background gravitational waves.

It is convenient to define the smeared constraints ${\cal H}_L:= \int d^3x {\cal H}(x) L(x)$, and ${\cal H}(\overrightarrow{L}) =   \int d^3x {\cal H}_i(x) L^i(x)$, for scalar functions $L$ and vector fields ${\cal L}_i$ on $\Sigma$. The constraints satisfy the Dirac algebra,
\begin{eqnarray}
\{{\cal H}(L), {\cal H}(L')\} &=& {\cal H}(\overrightarrow{L_c}) \\
\{{\cal H}(\overrightarrow{L}), {\cal H}(L')\} &=&  {\cal H}({\cal L}_{\overrightarrow{L}}L')\\
\{{\cal H}(\overrightarrow{L}), {\cal H}(\overrightarrow{L}')\} &=& {\cal H}([\overrightarrow{L}, \overrightarrow{L}']),
\end{eqnarray}
where $L_c^i = h^{ij}(L\partial_j L' - L' \partial_jL)$.
Note that ${\cal H}(\overrightarrow{L})$ is the generator of spatial diffeomorphisms. For any local observable $A(x)$, $\{{\cal H}(\overrightarrow{L}), A(x)\} = {\cal L}_{\overrightarrow{L}}A(x)$.

As a result of the symmetry diffeomorphism invariance, GR is a parameterised system, in the sense that its Hamiltonian vanishes due to the presence of first-class constraints\footnote{In this paper, we use the standard terms "parameterised system" and "parameterised field theory", which have   different meanings. A parameterised system is a system with a Hamiltonian that vanishes due to first-class constraints. A parameterised field theory is a field theory in which the spacetime coordinates associated to a foliation have been promoted to dynamical variables. A parameterised field theory is a parameterised system, the converse does not hold in general. For example, GR is a parameterised system, but not a parameterised field theory.}.
We must note a crucial difference between constrained systems of the Yang-Mills type and parameterised theories like GR. For gauge theories, the changes generated by the constraints do not change the physical state of the system.  By contrast, in parameterised theories the changes induced by the constraints are those associated with the dynamical evolution of the system \cite{ADM}. The true physical degrees of freedom are moved along the dynamical path. The different roles between the constraints resulting from gauge invariance and those resulting from reparameterisation invariance   have fundamental consequences for the quantum theory. This is  due to the fact that  in quantum theory, time is  clearly distinguished from all other physical variables and cannot be represented by a self-adjoint operator.

In order to identify the true degrees of freedom in a system with first-class constraints \cite{HeTe}, we must first identify the constraint surface $C$, i.e., the submanifold of the  phase space in which the constraints are satisfied. Then, we identify the reduced phase space as the quotient $C/\sim$, where the equivalence relation $\sim$ is defined as follows: two points $p$ and $p'$ of $C$ are equivalent if they are related by a gauge transformation generated by the constraints.

\subsection{Linearisation}
Next, we proceed to an analysis of the Hamiltonian around Minkowski spacetime, with no matter ($h_{ij} = \delta_{ij}, N = 1, N^i = 0, \pi^{ij} = 0, q_a = 0, p^a=0$). We introduce the formal expansion parameter $\lambda$, and we expand
\begin{eqnarray}
 h_{ij} = \delta_{ij}+ \lambda  \gamma_{ij},\hspace{0.3cm}
  N = 1 + \lambda n, \hspace{0.3cm}
  N_i = \lambda n_i, \hspace{0.3cm}
  \pi^{ij} \rightarrow \lambda \pi^{ij}. \label{linerr}
\end{eqnarray}
This expansion recovers the results of the Lagrangian linearisation (\ref{lagrangianlin}). It leads to a description for the linearised gravitational field interacting with matter that is similar to the  electromagnetic theory. The expansion (\ref{linerr})  is fundamentally problematic, because it incorporates {\em partial gauge-fixing} in the very first step: it determines the lowest order form of $N$ and $N_i$, which are arbitrary functions. The resulting system is not  a genuine parameterised system, so the symmetry of the full theory is lost. This is not a problem in classical theory, but it leads to strong ambiguities in any quantum description.

We will proceed with the linearisation (\ref{linerr}), because of the simplicity of the resulting descriptions, and then, we will present a conceptually more satisfying procedure in Sec. 3. Note that the parameter $\lambda$ is of the same status as the perturbation parameter in usual textbooks of perturbation theory. It is used in order to distinguish between  the different orders of the expansion, and after the orders have been separated, it is  set to unity. Physically, a perturbation analysis is meaningful if the system manifests a small dimensionless parameter. Typically this dimensionless parameter is proportional to $\kappa$ and the square of  some mass parameter that characterizes the matter distribution.

For the matter degrees of freedom, we substitute ${\cal V}$ and ${\cal V}_i$ with $\lambda {\cal V}$ and $\lambda {\cal V}_i$. We keep terms up to $\lambda^2$ in the Hamiltonian
\begin{eqnarray}
H = \lambda \int d^3x {\cal E} + \lambda^2 \int d^3x \left[\kappa(\pi^{ij}\pi_{ij} - \frac{1}{2} \pi^2) + \kappa^{-1} V(\gamma)  \right. \nonumber \\
\left. +\kappa^{-1} n [\partial^2\gamma - \partial_i \partial_j
\gamma^{ij} + \kappa {\cal E} ] +n_i(-2
\partial_j \pi^{ji} + {\cal P}^i) + \gamma_{ij} I^{ij}
\right] \label{hamillin}
\end{eqnarray}
where $\gamma = \gamma_{ij}\delta^{ij}$,
\begin{eqnarray}
{\cal E}(q, p) := {\cal V}(\delta_{ij}, q, p), \hspace{0.3cm}
{\cal P}_i(q, p) := {\cal V}_i(\delta_{ij}, q, p), \hspace{0.3cm}
I^{ij}(q, p) := \frac{\partial {\cal V}}{\partial h_{ij}}(\delta_{ij}, q, p),
\end{eqnarray}
and
\begin{eqnarray}
V := - \frac{1}{2} \partial_k\gamma_{ij} \partial^i \gamma^{kj} -   \frac{1}{4} \partial_k \gamma \partial^k \gamma + \frac{1}{2} \partial_i \gamma \partial_k \gamma^{ik} + \frac{1}{4} \partial_k \gamma_{ij} \partial^k \gamma^{ij}.
\end{eqnarray}
 Now the system is characterised by the constraints
 \begin{eqnarray}
 {\cal C} &=& \kappa^{-1}(\partial^2\gamma - \partial_i \partial_j
\gamma^{ij}) +  {\cal E} = 0 \label{const1a}\\
{\cal C}_i &=& -2
\partial_j \pi^{ji} + {\cal P}^i = 0, \label{const2a}
 \end{eqnarray}
which coincide with the  superhamiltonian and supermomentum constraints to first order in $\lambda$.

We complete linearisation by setting $\lambda = 1$. Hence, we obtained
 a new constrained system with symplectic form
\begin{eqnarray}
\Omega = \int d^3 x [ \delta \pi^{ij}(x) \wedge \delta \gamma_{ij}(x) + \delta p^a(x) \wedge \delta q_a(x)],
\end{eqnarray}
and Hamiltonian given by
\begin{eqnarray}
H =  \int d^3x \left[ {\cal E}  + \kappa(\pi^{ij}\pi_{ij} - \frac{1}{2} \pi^2) + \kappa^{-1} V(\gamma)  + \gamma_{ij} I^{ij} +  n {\cal C}
+n_i  {\cal C}^i \right]. \label{hamillin2}
\end{eqnarray}
The system is no more parameterised. The gauge choice implicit in the linearisation has selected the background coordinate $t$ as the parameter of Hamiltonian time evolution.

\subsection{The reduced state space}

Next, we split the metric perturbation into components, in a way that simplifies the constraint equations. We express the Fourier transform $\tilde{\gamma}_{ij}({\pmb k})$ as
\begin{eqnarray}
\tilde{\gamma}_{ij} = -i\kappa (k_i \tilde{\chi}_j + k_j \tilde{\chi}_i) + \sqrt{2 \kappa} \tilde{w}_{ij} + \frac{1}{2} \tilde{\phi} \Pi_{ij}, \label{split}
\end{eqnarray}
where $\Pi_{ij} := \delta_{ij} - \frac{k_ik_k}{|{\bf k}|^2} $ is the projector into the transverse subspace of $\tilde{\gamma}_{ij}$, $\tilde{w}_{ij}$ is the transverse-traceless (TT) component of $\tilde{\gamma}_{ij}$ ($\Pi^{ik} \Pi^{jl}w_{kl} = w^{ij}, \Pi^{ij}w_{ij} = 0$), and $\phi := \Pi^{ij}\gamma_{ij}$. Note that the uneven dependence of the components of $\gamma_{ij}$ on $\kappa$ does not affect physics, because $\kappa$ is not an expansion parameter. We have chosen this expression in order to simplify the correspondence of our results with the Newtonian description.

We substitute Eq. (\ref{split}) into the gravitational component of the symplectic potential $\Theta_g = \int d^3 x \pi^{ij} \delta \gamma_{ij} = \int \frac{d^3k}{(2\pi)^3} \tilde{\pi}^{ij} \delta \tilde{\gamma}_{ij}$, to obtain
\begin{eqnarray}
\Theta_g = \int \frac{d^3k}{(2\pi)^3} \left[(-2i\kappa k_i\tilde{\pi}^{ij})\delta \tilde{\chi}_j + \sqrt{2 \kappa} \tilde{\pi}^{kl} (\Pi_k^i\Pi_k^j - \Pi^{ij}\Pi_{kl})\delta \tilde{w}_{ij} + \frac{1}{2}\tilde{\pi}^{ij} \Pi_{ij}\delta \tilde{\phi}\right]
\end{eqnarray}
Hence, we identify the Fourier conjugate variables $\tilde{\pi}_\chi^{i}:= -2i\kappa k_i\tilde{\pi}^{ij}, \tilde{\pi}_{w}^{ij} := \sqrt{2 \kappa}\tilde{\pi}^{kl} (\Pi_k^i\Pi_l^j - \Pi^{ij}\Pi_{kl})$, and $\tilde{\pi}_{\phi} :=  \frac{1}{2}\tilde{\pi}^{ij} \Pi_{ij}$ to $\tilde{\chi}_i, \tilde{w}_{ij}$ and $\tilde{\phi}$, respectively. We Fourier-transform back to obtain
 \begin{eqnarray}
\Theta_g = \int d^3x [\pi_{\chi}^{j}(x)\delta \chi_j(x) + \pi^{ij}_{w}(x) \delta w_{ij}(x) + \pi_{\phi}(x) \delta \phi(x)].
\end{eqnarray}
The constraints simplify when expressed in terms of the new variables
 \begin{eqnarray}
{\cal C} = \kappa^{-1}\nabla^2 \phi + {\cal E} = 0 \label{constraint1}\\
{\cal C}^i = \kappa^{-1} \pi_{\chi}^i +{\cal P}^i = 0.
\end{eqnarray}
We also define the smeared constraints ${\cal C}(L) = \int d^3x L(x) {\cal C}(x)$ and ${\cal C}(\overrightarrow{L}) = \int d^3x L_i(x) {\cal C}^i(x)$

Next, we evaluate the various terms of the Hamiltonian on the constraint surface
\bey
\pi^{ij} \pi_{ij} - \frac{1}{2} \pi^2 = \frac{1}{2\kappa} \pi_w^{ij} \pi_{w ij} + \frac{1}{2\kappa^2} \pi_\chi^i \Delta_{ij} \pi_\chi^j + \frac{1}{\kappa} \pi_\phi \nabla^{-2} \partial_i \pi_\chi^i + \partial_i R^i, \\
 V(\gamma) = \frac{\kappa}{2}  \partial_kw_{ij} \partial^k w^{ij} -\frac{1}{8}  (\nabla \phi)^2   + \kappa^2 \partial_iK^i
\eey
where
\bey
 R^i &=& \frac{\sqrt{2}}{\kappa^{3/2}} \pi_w^{ij} \widetilde{\Delta}_{jk} \pi_\chi^k -\frac{\sqrt{2}}{\sqrt{\kappa}} \pi_w^{ij}\partial_j\nabla^{-2}\pi_\phi + \frac{2}{\kappa} \Pi^{ij} \pi_\phi \widetilde{\Delta}_{jk} \pi_\chi^k - \Pi^{ij} \pi_\phi \partial_j \nabla^{-2} \pi_\phi  \nonumber\\
 & &+\frac{1}{2\kappa^2} \widetilde{\Delta}_{jk} \pi_\chi^k \partial^j \widetilde{\Delta}^i_n \pi_\chi^n - \frac{1}{4\kappa^2} \widetilde{\Delta}^i_j \pi_\chi^j \partial_k \nabla^{-2} \pi_\chi^k + \frac{1}{2\kappa^2} \partial^i \nabla^{-2} \pi_\chi^j \nabla^{-2} \Pi_{jk} \pi_\chi^k \nonumber \\
 & & + \frac{1}{8\kappa^2} \partial_j \nabla^{-4} \pi_\chi^j \Pi^i_k \pi_\chi^k + \frac{1}{8\kappa^2} \partial^i \partial^j \partial_k \nabla^{-4} \pi_\chi^k \partial_j \partial_n \nabla^{-4} \pi_\chi^n
 \eey
 \bey
K^i = \partial_j \chi^i \nabla^2 \chi^j - \partial_j \chi^k \partial^i \partial_k \chi_j - \frac{1}{2 \kappa} \phi \left( \nabla^2 \chi^i - \partial^i \partial_j \chi^j \right) - \frac{1}{\kappa} w_{jk} \partial^k w^{ji} \\
+ \frac{1}{\kappa^2} \partial_j \partial_k \nabla^{-2} \phi \partial^i \partial^j \partial^k \nabla^{-2} \phi - \frac{1}{\kappa^2} \partial_j \phi \partial^i \partial^j \nabla^{-2} \phi,
\eey
\bey
\Delta_{ij} = - \nabla^{-2} \left( \delta_{ij} - \frac{3}{4} \partial_i \partial_j \nabla^{-2} \right) \\
\widetilde{\Delta}_{ij} = - \nabla^{-2} \left( \delta_{ij} - \frac{1}{2} \partial_i \partial_j \nabla^{-2} \right) \\
\Pi_{ij} = \delta_{ij} - \partial_i \partial_j \nabla^{-2},
\eey
and
\bey
\nabla^{-2} f(x) = -  \int d^3x' \frac{f({\bf x}')}{4\pi|{\bf x} - {\bf x}'|} .
\eey

The constraint surface $C$ is spanned by the variables $w_{ij}, p_w^{ij}, \chi_i, \pi_{\phi}, q_a, p^a$. The TT components $w_{ij}, p_w^{ij}$ commute with the constraints, hence, they are true degrees of freedom. The vector field $\chi_i(x)$ satisfies
\bey
\{\chi_i(x) ,  {\cal C}(L) + {\cal C}(\overrightarrow{L})\} =  \kappa^{-1} L_i(x), \label{chill}
\eey
i.e., it is essentially a parameter of the gauge orbit generated by the supermomentum constraint, i.e., spatial diffeomorphisms.

Similarly, we find
\begin{eqnarray}
\{ \pi_\phi (x), \mathcal{C}(L) + \mathcal{C} (L) \} = - \kappa^{-1} \nabla^2 L(x)
\end{eqnarray}
This implies that the function
\begin{eqnarray}
\tau (x) = - \kappa \nabla^{-2} \pi_\phi (x)
\end{eqnarray}
satisfies
\begin{eqnarray}
\{\tau(x),          {\cal C}(L) + {\cal C}(\overrightarrow{L})  \} =  L(x), \label{tll}
\end{eqnarray}
i.e., $\tau(x)$ is a coordinate of the gauge orbit generated by the superhamiltonian constraint.

To understand the physical meaning of the quantities $\tau$ and $\chi_i$, we note that for the matter variables
\bey
\{q_a(x),  {\cal C}(L) + {\cal C}(\overrightarrow{L}) \} = \{q_a(x), H(L, \overrightarrow{L}) \}\\
\{p^a(x),  {\cal C}(L) + {\cal C}(\overrightarrow{L}) \} = \{p^a(x), H(L, \overrightarrow{L}) \}
\eey
where $H(L, \overrightarrow{L}) = \int d^3x [L(x){\cal E}(x) + L_i{\cal P}^i(x)]$ is the Hamiltonian for a parameterised relativistic scalar field in flat spacetime. It corresponds to evolution of the scalar field  along arbitrary foliations of the flat spacetime, each choice of foliation being in correspondence with a choice of $L$ and $L_i$.

Let $Q_a(x, t)$ be a solution to the field equations along a Lorentzian foliation; this solution is a functional of $q_a, p^a$, which define the initial conditions at $t = 0$. Consider a foliation with time-coordinate defined by the surface $t:=\tau(x)$ and space coordinates $y_i := \chi^i(x)$, and express the same solution as a functional $Q_a(t, y; \tau, \chi, q, p)$.
 By Eqs. (\ref{chill}, \ref{tll}), the action of the constraints on $q_a, p^a$ (evolution along the foliation) compensates on the action on $\tau, \chi$, so that the  functional $Q_a(t, y; \tau, \chi, q, p)$ commutes with the constraints. Hence, the true degrees of freedom for matter correspond to solutions of the equations of motion for matter, modulo all possible parameterisations of spacetime.

\subsection{Gauge fixing}

The meaning of the quantities $\pi_{\phi}$ and ${\cal \chi}_i$ as parameters specifying the choice of space-time coordinates
was discovered by Arnowitt, Deser, and Misner (ADM) \cite{ADM2}. In
an effort to
identify the proper dynamical degrees of freedom of
the gravitational field, they proposed a gauge-fixing condition
\bey
t = \tau(x), \hspace{0.5cm}  \chi_i(x) = 0 \label{ADMgauge}
\eey
that defines a coordinate system (a spacelike foliation) that is  as close as possible
to a Cartesian system of coordinates in the flat
space-time. Hamilton's equation of motion for $\pi^{ij}$ and $\gamma_{ij}$ with this condition imply that $n = n_i = 0$.

Eq. $t = \tau(x)$ defines the spacelike surfaces $\Sigma_t$ of constant $t$, in a spacetime described by coordinates $(t, x^i)$. On the other hand $\chi^i(x)$ defines to  leading order a deformation of the (flat) spatial coordinates by $x^i \rightarrow x^i + \chi^i(x)$, due to the presence of the gravitational field. Hence, the  ADM gauge implies the use of the background spatial coordinate system.

In the ADM gauge, the Hamiltonian becomes a sum of three terms
\bey
H_{ADM} = \int d^3x  \left[ {\cal E}(x) +   {\cal H}_{gsi}(q, p)+{\cal H}_{GW}(w, \pi_w, q, p) \right], \label{hadm}
\eey
where
\bey
{\cal H}_{gsi} = \frac{\kappa}{2} \left[ -\frac{1}{4} \partial_i \nabla^{-2} \mathcal{E} \partial^i \nabla^{-2} \mathcal{E} + \mathcal{P}^i \Delta_{ij} \mathcal{P}^j - \Pi_{ij} \nabla^{-2} \mathcal{E} I^{ij} \right],
\eey
is the Hamiltonian density of gravitational self-interaction, and
\bey
{\cal H}_{GW} =   \frac{1}{2}  \pi_w^{ij} \pi_{wij} +  \frac{1}{2}  \partial_k w_{ij}\partial^k w^{ij} + \sqrt{2 \kappa} w_{ij}I^{ij}
\eey
 is the Hamiltonian density for gravitational waves, including a term for the interaction of gravitational waves with matter.

In the ADM gauge, the lapse vector $N = 1$ and the shift vector $N_i = 0$. However, the expression for the three metric is rather complex. Even for $w_{ij} = 0$,
so that the spacetime metric is a non-local functional of the Newtonian potential  $\phi(x) = -\kappa \nabla^{-2}{\cal E}$,
\bey
ds^2 = - dt^2 + dx^i dx^j  \left[\delta_{ij} + \frac{1}{2} \Pi_{ij}\phi \right]. \label{metricADM}
\eey
The three metric is simpler in the {\em isotropic gauge}, determined by
\bey
\chi_i - \frac{1}{4} \nabla^{-4} \partial_i {\cal E} = 0,  \hspace{0.3cm}  \tau + \frac{\kappa}{4} \nabla^{-4}\nabla_i {\cal P}^i = t.
\eey
In the isotropic gauge, the three-metric for $w_{ij} = 0$ is
\bey
h_{ij} = \delta_{ij} \left(1 - \frac{1}{2} \phi \right).
\eey

\section{Weak gravity as a parameterised field theory}

The problem with the linearisation expansion of Sec. 2.2 is that it involves a partial gauge fixing, as it fixes  $N$ and $N_i$ to lowest order. The time-reparameterisation symmetry of the initial Hamiltonian is thereby lost. The description of Sec. 2.2 is what one obtains when taking the linearisation at the Lagrangian level and then Legendre-transforming to a Hamiltonian description. This procedure breaks general covariance and misrepresents the causal structure of the system.
In this section, we present an alternative approach to  linearised gravity that preserves time-reparameterisation symmetry. This approach is an adaptation of the classic Quantum Geometrodynamics analysis of gravitons by Kuchar \cite{Kuchar70}.

\subsection{Perturbation expansion of the constraints}

The key idea in this expansion scheme is that we expand the constraints (\ref{superH}, \ref{superM}) with respect to the perturbations of the metric and the momentum around the phase space point $(\delta_{ij}, 0)$. We do not expand the lapse and a shift variables around a background value. Thus, we avoid partial gauge fixing, and the Hamiltonian still vanishes due to the first-class constraints. The difference is that we have to keep terms to second order to the expansion in $\lambda$ in order to obtain meaningful weak-field dynamics.

We write $h_{ij} = \delta_{ij} + \lambda \gamma_{ij}$, $\pi^{ij} \rightarrow \lambda \pi^{ij}$, where $\lambda$ is an expansion parameter. We keep terms up to the first two orders in $\lambda$, to obtain
 \bey
{\cal H} &=& \lambda  \left[ \kappa^{-1}(\partial^2\gamma - \partial_i \partial_j
\gamma^{ij})
+  {\cal E}\right]
\nonumber \\
&+& \lambda^2 \left[ \kappa(\pi^{ij}\pi_{ij} - \frac{1}{2} \pi^2) + \kappa^{-1} V(\gamma)  + \gamma_{ij} I^{ij} + \kappa^{-1} \partial_i J^i(\gamma) \right], \label{constr1}
\\
{\cal H}^i &=& \lambda \left( -2
\partial_j \pi^{ji} + {\cal P}^i \right)+ \lambda^2 \left[ L^i + M^{ijk} \gamma_{jk}\right], \label{constr2}
\end{eqnarray}
where
\bey
L^i = -\pi^{ij} \partial_j\gamma -2\partial^k \left( \gamma^{ij} \pi_{jk} \right) + \pi^{jk} \partial^i \gamma_{jk},
\eey
 $M^{ijk} = \partial {\cal V}^i/\partial h_{ij}$, and
\bey
J^i = \frac{1}{2} \gamma \partial^i\gamma + \frac{1}{2} \gamma \partial_j \gamma^{ij}  -\gamma^{jk}\partial^i\gamma_{jk}   + \partial_j(\gamma^{jk}\gamma^{i}{}_k -\gamma \gamma^{ij}).
\eey

We make a canonical transformation from $\phi, \pi_{\phi}$ to $  \tau = - \kappa \nabla^{-2} \pi_\phi$ and $ \pi_\tau = \kappa^{-1} \nabla^2 \phi$. By Eqs. (\ref{constr1}---\ref{constr2}), $ \pi_\tau + \mathcal{E} = O(\lambda) $, and $\pi_{\chi}^i + \kappa {\cal P}^i = O(\lambda)$, so we can remove $\pi_{\tau}$ and $\pi_{\chi}^i$ from the $\lambda^2$ terms in Eqs. (\ref{constr1}, \ref{constr2}) up to terms of order $\lambda^3$.

Hence, up to terms of order $\lambda^3$, the two constraints can be solved for $\pi_{\tau}$ and $\pi_{\chi}^i$, as
\begin{eqnarray}
\pi_{\tau} + {\cal G}^0(\chi, \tau, w, \pi_w, q, p) = 0\\
\pi_{\chi}^i + {\cal G}^i(\chi, \tau, w, \pi_w, q, p) = 0
\end{eqnarray}
in terms of the functions
\begin{eqnarray}
 {\cal G}^0(\chi, \tau, w, \pi_w, q, p) = {\cal E}+ {\cal H}_{gsi}+{\cal H}_{GW}
 - 2 \kappa \partial_i \chi_j I^{ij} + \nabla^2 \tau \nabla^{-2}\partial_i{\cal P}^i + \kappa^{-1} \partial_i \bar{J}^i
 \\
{\cal G}^i(\chi, \tau, w, \pi_w, q, p) = \kappa {\cal P}^i + \kappa L^i + \sqrt{2} \kappa^{3/2} M^{ijk} w_{jk} - 2 \kappa^2 M^{ijk} \partial_j \chi_k - \kappa^2 M^{ijk} \Pi_{jk} \nabla^{-2}{\cal E}
\end{eqnarray}
where
\bey
\bar{J}^i = J^i +\kappa R^i +\kappa^2 K^i.
\eey

The symplectic potential on the constraint surface is
\begin{eqnarray}
\Theta = \int d^3x \left[ p^a(x)\delta q_a(x) + \pi_w^{ij}(x)\delta w_{ij}(x) - {\cal G}^0 \delta \tau -{\cal G}_i \delta \chi^i\right]. \label{symplecticpot}
\end{eqnarray}

Comparing with Eq. (\ref{parameterized}), we see that the symplectic potential is that of a parameterised field theory. A PFT is obtained from an ordinary field theory in a background spacetime by promoting the spacetime coordinates $X^{\alpha}$ to dynamical fields---see, \cite{Lan49, Kuchar88, Kuchar89} and also the Appendix. The resulting Hamiltonian system has first-class constraints of the form
\bey
{\cal C}^{\alpha} := P_X^{\alpha} + f^{\alpha}(\phi, \pi_{\phi}) = 0, \label{constrpft}
\eey
 where $P_X^{\alpha}$ is the conjugate momentum to $X^{\alpha}$, and $\phi$ and $\pi_{\phi}$ are the field and its conjugate momentum. The Hamiltonian of the PFT vanishes because of the constraints, so the system is parameterised. In PFTs, the symplectic potential on the constraint surface is of the form
\begin{eqnarray}
\Theta = \int d^3x \left[ \pi_\phi(x) \delta \phi(x)  - f_{\alpha} (\phi, \pi_{\phi}) \delta X^{a}(x) \right],
\end{eqnarray}
i.e., Eq. (\ref{symplecticpot}).

Ever since the first works on canonical quantisation \cite{Dirac2, ADM}, it has been suggested that GR can be expressed as a PFT, in the sense that it would be possible to find a canonical transformation through which the constraints would take the form (\ref{constrpft}). This would provide strong impetus towards its canonical quantisation, by `parameterising' the theory.  However, as pointed out in Ref. \cite{Kuchar92}, the full GR constraints are quadratic in momentum, and hence, it is implausible that they can be brought in the form (\ref{constrpft}) in the full phase space---see, also Ref. \cite{Torre}.

 Linearised GR with matter has constraints of the form (\ref{constrpft}), and as such it is equivalent to a PFT. This means that in weak gravity, we can directly apply methods for quantising PFTs to the system, and, in principle, test their predictions against experiments on macroscopic quantum systems.

\subsection{Deparameterisation through gauge fixing}
There are two different ways to describe a PFT. The simplest way is deparameterise the system. The route from an ordinary field theory to a PFT is unique, however, the converse does not hold. There is an infinity of different ways of parameterising a PFT and taking an ordinary fields theory on a fixed background.

Deparameterisation is essentially a form of gauge fixing.
 We can set $\tau$ and $\chi^i$ equal to specific functions of $x$ and $t$, say, $T_t(x)$ and $X^i_t(x)$. The surfaces of constant $t$ are defined by the condition $\tau = T_t(x)$, while the spatial coordinates at each $\Sigma_t$ are given by $x^i + X^i_t(x)$.

Then, the action for the reduced system
is
\bey
S[q,p, w, \pi_w] = \int dt d^3 x \left[\pi_w^{ij}\dot{w}_{ij} + p^a\dot{q}_a - \dot{T}_t(x)  {\cal G}^0 -\dot{X}^i_t {\cal G}_i\right]. \label{action}
\eey

The associated Hamiltonian is
\bey
H = \int d^3 x \left[\dot{T}_t(x)  {\cal G}^0  + \dot{X}^i_t {\cal G}_i\right]. \label{hamtsont}
\eey
Physical predictions should be the same irrespective of gauge fixing. Each deparameterisation leads to a different spacetime picture for the solutions to the equations of motion. However, physics is not affected. At least in classical GR, solutions  that correspond to different deparameterisations must be related by  a spacetime diffeomorphism.

Note that if we choose $X_t(x)$ to be $t$-independent, then only ${\cal G}^0$ contributes in the action. If also $T_t(x) = t$, then the Hamiltonian (\ref{hamtsont}) coincides with the ADM Hamiltonian.

For $T_t(x) = a_0 t + f_0(x)$ and $X^i_t(x) = a^it + f^i(x)$, the Hamiltonian is
\begin{eqnarray}
H = a_0 \int d^3x {\cal G}^0 + a_i \int d^3x {\cal G}^i,
\end{eqnarray}
and the boundary terms in $G^0$ and $G^i$ drop out.

\subsection{The bubble-time formalism}

If we want to avoid gauge fixing, we must take into account that the true degrees of freedom carry an explicit functional dependence on the foliation variables  ${\cal X}^{\alpha} = (\tau, \chi^i)$, which determine a spacelike foliation.
Here, we introduce the index $\alpha = (0, i)$. We will also write ${\cal G}^{\alpha} = ({\cal G}^0, {\cal G}^i)$.
It is important to emphasize that the variables ${\cal X}^{\alpha}$ are not coordinates but foliations, i.e., they are to be identified with spacetime diffeomorphisms.

The true degrees of freedom  $q, p, w$ and $\pi_w$ are then written as $q_a(x,{\cal X}]$, $p^a(x, {\cal X}]$, $w_{ij}(x, {\cal X}]$, $\pi_w^{ij}(x, {\cal X}]$. These quantities satisfy the functional evolution equations
\begin{eqnarray}
\frac{\partial q_a(x, {\cal X}) }{\partial {\cal X}^{\alpha}(x')} = \{ q_a(x, {\cal X}], {\cal G}^{\alpha}(x', {\cal X}]\}  \\
\frac{\partial p^a(x, {\cal X} )}{\partial {\cal X}^{\alpha}(x')} = \{ p^A(x, {\cal X}], {\cal G}^{\alpha}(x', {\cal X}]\}  \\
\frac{\partial w_{ij}(x, {\cal X}) }{\partial {\cal X}^{\alpha}(x')} = \{ w_{ij}(x, {\cal X}], {\cal G}^{\alpha}(x', {\cal X}]\}  \\
\frac{\partial \pi_w^{ij}(x, {\cal X}) }{\partial {\cal X}^{\alpha}(x')} = \{ \pi_w^{ij}(x, {\cal X}], {\cal G}^{\alpha}(x', {\cal X}]\}.
\end{eqnarray}
The Poisson bracket above is defined on the reduced state space, i.e., with respect only to the $q, p, w, \pi_w$ degrees of freedom.

This is the so-called ‘bubble-time’ or ‘multi-time’ canonical formalism \cite{Kuchar72}. The time evolution of the true degrees of freedom is given with respect to any spacelike foliation, thus, manifesting explicitly the coordinate independence of GR. The gauge-fixed description carries less information than the bubble-time description, in the sense that it expresses physics only with respect to a given foliation. Of course, in classical GR, physics is the same in all foliations. This is not the case when we quantise the theory.

PFTs can also be treated with more standard Hamiltonian methods, in a way that also allows one to take boundary conditions into account, see, for example. Refs. \cite{Barbero1, Barbero2}.

\subsection{The Newtonian regime}
The action (\ref{action}) contains complete information about the interaction of weak gravity with matter.  The transverse-traceless part that corresponds to gravitational waves is insignificant, if no gravitational waves exist at the initial moment of time and if the density and acceleration of matter is sufficiently low.

The Newtonian regime  is obtained by keeping only the terms of energy density for matter, and dropping all terms that involve ${\cal P}^i, I^{ij}$ and $M^{ijk}$. Then,

\begin{eqnarray}
 {\cal G}^0(\chi, \tau, w, \pi_w, q, p) &=& {\cal E} -\frac{\kappa}{8} \partial_i \nabla^{-2}{\cal E}\partial^i \nabla^{-2}{\cal E}  + \kappa^{-1} \partial_i \bar{J}^i\label{G0N}\\
{\cal G}^i(\chi, \tau, w, \pi_w, q, p) &=& \kappa L^i.
\end{eqnarray}

In  the ADM   gauge,
\bey
H = \int d^3x {\cal E}(x) - \frac{G}{2} \int d^3x d^3x' \frac{{\cal E}(x){\cal E}(x')}{|{\bf x} - {\bf x}'|}, \label{intham}
\eey
as expected in Newtonian gravity.

\section{Effective QFT description}

\subsection{Quantisation methods}
There are two major approaches in canonically quantising a constrained system, the reduced state space quantisation and the Dirac quantisation. In the former, the constraints are implemented before quantisation, while in the latter the constraints are implemented after quantisation. Both approaches end up with the construction of a Hilbert space for the quantised true degrees of freedom and the associated observables.

There are also two major path-integral approaches to quantisation. In Feynman path integrals \cite{Fadeev, FaVi, BaVi}, one works at the level of the unphysical variables (i.e., prior to the implementation of the constraints) and the emphasis is on constructing a consistent integration measure that leads to unambiguous physical predictions. Whenever the Hilbert space of true observables is constructed, the analysis is closer to  Dirac quantization \cite{HoVo}.
  In coherent-state path integrals \cite{Klauder}, the emphasis is on the construction of kernel that defines the Hilbert space of the true degrees of freedom. Eventually it reduces to a form of Dirac quantisation.  Note that path integrals for parameterised theories are very different from those for other theories with constraints, like Yang Mills \cite{KuHa1, KuHa2}.

In this paper, we will only consider canonical quantisation methods. These suffice in order to clarify our main points. However, path integral studies of weak gravity may prove helpful in further explicating the symmetries of the  theory, and as such they are an important follow-up to this paper.

\subsection{Quantisation in the ADM gauge}
In general, reduced-state space quantisation is the physically most transparent method, but also the most difficult to implement  technically. It is transparent, because it is not concerned with the quantum representation of pure-gauge degrees of freedom. It is technically demanding  because the reduced state space is a quotient space that may not have a manifold structure. There are some procedures for quantising even such systems; however,  the degree of difficulty is such as to discourage the attempt of all, but the simplest cases.

Weak gravity interacting with matter is one of these cases, at least, in some gauges. In the ADM gauge, quantisation leads straightforwardly to an effective QFT. The fundamental variables are local fields $q_a(x), p^a(x), w_{ij}(x), \pi_w^{ij}(x)$, that enter the Hamiltonian (\ref{hadm}). This system can be treated by standard perturbative QFT quantisation.

The unperturbed system consists of the matter fields $\hat{q}_a(x), \hat{p}^a(x)$ evolving under the Hamiltonian $\hat{H} = \int d^3x {\cal E}([\hat{q}(x), \hat{p}(x)]$ in the Hilbert space ${\cal H}_{mat}$ and the gravitons under the Hamiltonian
\bey
\hat{H}_{gr} := \int d^3x \left[\frac{1}{2}  \hat{\pi}_w^{ij} \hat{\pi}_{wij} +  \frac{1}{2}  \partial_k \hat{w}_{ij}\partial^k \hat{w}^{ij} \right],
\eey
in the Hilbert space ${\cal H}_{grav}$.
The unperturbed system is then described by the Hilbert space ${\cal H}_{mat} \otimes {\cal H}_{grav}$.

There are two interaction terms. The term $\hat{w}_{ij} \hat{I}^{ij}$ generates the coupling of gravitons to matter and it is consistent up to  tree level. The second term is a non-local self-interaction term ${\cal H}_{gsi}$ that corresponds to the gravitational self-interaction of matter. It involves components of the stress-energy tensor that are not well defined as operators in standard QFT\footnote{Components of the stress-energy tensor are well defined only when integrated with a {\em spacetime} test function, not with a spatial smearing function.}. They require appropriate regularisation.

Here, we are mostly interested in the Newtonian regime, i.e., when the graviton field decouples from matter and matter can be treated using non-relativistic QFT. For concreteness, assume the matter fields to be fermions. In the non-relativistic limit, they are described by field operators $\hat{\psi}_a(x)$ and $\hat{\psi}_a^{\dagger}(y)$, such that
\bey
\{\hat{\psi}_a({\bf x}), \hat{\psi}^{\dagger}_b({\bf x}') \} = \delta_{ab}  \delta^3({\bf x}, {\bf x}').
\eey
The non-relativistic fields can be expressed in terms of fermionic creation and annihilation operators,
\bey
\hat{\psi}_a({\bf x}) = \int \frac{d^3p}{(2\pi)^3} e^{i{\pmb p} \cdot{\pmb x}} \hat{c}_a({\pmb p}), \hspace{0.5cm}\hat{\psi}^{\dagger}_a({\bf x}) = \int \frac{d^3p}{(2\pi)^3} e^{-i{\pmb p} \cdot{\pmb x}} \hat{c}_a^{\dagger}({\pmb p})
\eey
that satisfy the canonical anti-commutation relations.

Given the quantum fields above, we define the particle density functions
\bey
\hat{n}_a({\bf x})  = \hat{\psi}^{\dagger}_a({\bf x}) \hat{\psi}_a({\bf x}).
\eey
The matter Hamiltonian is
\bey
\hat{H}_{mat} = - \sum_a \int d^3x \frac{1}{2m_a}  \hat{\psi}^{\dagger}_a({\bf x})  \nabla^2 \hat{\psi}_a({\bf x}) + \frac{1}{2}\sum_{a \neq b}\int d^3x d^3x' V_{ab}({\bf x} - {\bf x'}) \hat{n}_a({\bf x}) \hat{n}_b({\bf x}),
\eey
where $m_a$ is the mass of a particle of type $a$ and $V_{ab}({\bf r})$ is a short-range interaction potential between the different matter particles.

The gravitational self-interaction Hamiltonian $\hat{H}_{gsi}$ is expressed in terms of the mass density
\bey
\hat{\mu} ({\bf x}) = \sum_a m_a \hat{n}_a({\bf x}), \label{massdens}
\eey
because in the non-relativistic regime the mass density is the dominant term in the energy density. Then,
\bey
\hat{H}_{gsi} = - \frac{G}{2} \int d^3x d^3x' \frac{\hat{\mu} ({\bf x})\hat{\mu} ({\bf x}')}{|{\bf x} - {\bf x}'|}. \label{hgsiq}
\eey
The Hamiltonian must be regularised through the introduction of a cut-off $\ell$ in the denominator so that $|{\bf x} - {\bf x}'|$ is substituted by $\sqrt{\ell^2 +|{\bf x} - {\bf x}'|^2}$. Then, there is a self-energy contribution $-\frac{Gm_a^2}{2\epsilon}$ to each particle that is absorbed into a mass renormalisation of the Hamiltonian.

This expression from the Hamiltonian can also be derived in non relativistic physics, by interpreting the gravitational interaction as an inter-particle potential.

\subsection{Bipartite systems}

The presence of an external constraints may allows us to express a subset of the matter Hilbert space into  a tensor product ${\cal H}_1 \otimes {\cal H}_2$, where ${\cal H}_1$ describes degrees of freedom localised in a region $C_1$ and ${\cal H}_2$ degrees of freedom in a region  $C_2$. This can be achieved by adding to the matter Hamiltonian an external potential $U(x)$ that vanishes only in $C_1$ and $C_2$ and takes very large values everywhere else. Hence, the subspace of sufficiently small eigenvalues of the Hamiltonian can be approximated by the Fock space for fields on $C_1 \times C_2$, hence, to be of the form ${\cal H}_1 \otimes {\cal H}_2$---see, Ref. \cite{AnHu20} for an example.
Then, we can write the number densities of each component as $\hat{\nu}_a^{(1)}$ and $\hat{\nu}_a^{(2)}$, and the associated mass densities  $\hat{\mu}^{(1)}$ and $\hat{\mu}^{(2)}$.

Assuming that the separation between the subsystems is much larger than the range of the interaction potential, the bipartite system is described by the Hamiltonian,
\bey
\hat{H} = \hat{H}_1 \otimes \hat{I} + \hat{I} \otimes \hat{H}_2 + \hat{H}_{int}, \label{Hamilbi}
\eey
where
\bey
\hat{H}_i &=&  - \sum_a \int_{C_i} d^3x \frac{1}{2m_a}  \hat{\psi}^{\dagger}_a({\bf x})  \nabla^2 \hat{\psi}_a({\bf x}) + \frac{1}{2}\sum_{a \neq b}\int d^3x d^3x' V_{ab}({\bf x} - {\bf x'}) \hat{n}_a^{(i)} ({\bf x}) \hat{n}^{(i)} _b({\bf x})\nonumber \\
 &-& \frac{G}{2}  \int_{C_i} d^3x \int_{C_i} d^3x' \frac{\hat{\mu}^{(i)} ({\bf x})\hat{\mu}^{(i)} ({\bf x}')}{|{\bf x} - {\bf x}'|} \\
\hat{H}_{int} =   &-&G \int_{C_1} d^3x  \int_{C_2} d^3x' \frac{\hat{\mu}^{(1)} ({\bf x})\hat{\mu}^{(2)} ({\bf x}')}{|{\bf x} - {\bf x}'|}. \label{hint}
\eey
The Hamiltonian (\ref{Hamilbi}) provides the foundation for any model about the gravitational interaction between separated quantum systems. It accounts for all set-ups that have been proposed so far, one simply has to project to the appropriate subspaces, in order to obtain a particle, or even a qubit description, see, for example, Ref. \cite{AnHu20} .

Obviously, the interaction Hamiltonian generates  entanglement between the subsystems. Its order of magnitude can be estimated perturbatively. Consider an eigenstate $|\psi_1\rangle \otimes|\psi_2\rangle$ of the Hamiltonian  $\hat{H}_1 \otimes \hat{I} + \hat{I} \otimes \hat{H}_2$ with energy $E$. The interaction term is of the order $\frac{Gm_1m_2}{L}$, where $m_i$ is the total mass of component $i$ and $L$ is the typical length scale of the system.  By elementary perturbation theory, the perturbed eigenstate has entanglement of the order of $\frac{Gm_1m_2}{LE}$.
The typical frequency of  Rabi oscillations, similar to the ones described in Ref. \cite{AnHu20} is $\frac{Gm_1m_2}{L}$.

\subsection{Spacetime properties}
In the Newtonian regime, there are no independent gravitational degrees of freedom. Nonetheless, we can
 define quantum metric observables. The idea is to interpret the classical equation for the metric (\ref{metricADM}) as a Heisenberg-type equation,
\bey
\hat{g}({\bf x}, t) = - dt^2 + dx^i dx^j \left[ \delta_{ij}+ 8 \pi G \Pi_{ij}\nabla^{-2} \hat{\mu}({\bf x}, t)\right], \label{gggg}
\eey
where $(t, x^i)$ are the coordinates in the ADM gauge, and $\hat{\mu}({\bf x}, t)$ is the Heisenberg-picture mass density operator (\ref{massdens}). Having a quantised metric does not mean that gravity is quantised, for after all, $\hat{g}(x, t)$ is an operator on the Hilbert space of matter degrees of freedom.

The quantum metric $\hat{g}$ is not itself an observable, rather it serves as a generator of many different observables.
To see this, consider  a classical test particle along a trajectory given by ${\bf x}(t)$ in the ADM frame, with initial point ${\bf x}_0$ at $t = 0$ and final point ${\bf x}_f$ at $t = T$.  For a metric $ds^2 = - dt^2+(\delta_{ij} + \gamma_{ij})dx^i dx^j$ where $\gamma_{ij}$ is a perturbation, the proper time of this trajectory is
\bey
\tau = \tau_0 -\frac{1}{2} \int_0^T dt \gamma_{ij}(x(t), t) \frac{\dot{x}^i \dot{x}^j}{1 - \dot{\bf x}^2} \label{tttt}
\eey
where $\tau_0 = \int_0^T dt \sqrt{1 - \dot{\bf x}^2}$. The corresponding quantum observable for Eq.  (\ref{tttt}) is
\bey
\hat{\tau} = \tau_0 \hat{I} - 4 \pi G  \int_0^T dt \frac{\dot{x}^i(t) \dot{x}^j(t)}{1 - \dot{\bf x}^2(t)} \Pi_{ij}\nabla^{-2} \hat{\mu}({\bf x}(t), t) , \label{tauobs}
\eey
The difference from the Minkowski proper time $\delta \hat{\tau} := \hat{\tau} - \tau_0 \hat{I} $ is  a time-smeared quantum observable. For the quantum measurement theory of such observables, see, \cite{AnSav19}. Having defined spacetime observables, we can also define quantum-controlled unitary operations, and there are arguments that this leads to entanglement of temporal order between different events \cite{ZCPB}.

There is a severe problem with such definitions: they are highly gauge-dependent. To see this, let us compare with the quantum description of the system in the isotropic gauge. Let ${\cal H}_{ADM}$ and ${\cal H}_{iso}$ be the Hilbert spaces of the quantum systems in the different gauges. Let $\xi^{\mu}(\cdot)$ be a spacetime path: in the ADM coordinates, it defines a path $x_{ADM}(t)$ and in the isotropic coordinates a path $x_{iso}(t)$. Hence, we can define a clock observable $\hat{\tau}_{ADM}$ in ${\cal H}_{ADM}$ and another clock observable $\hat{\tau}_{iso}$ in ${\cal H}_{iso}$.

The relation between the spacetime coordinates in the two gauges is
\bey
t_{iso} = t_{ADM} +\frac{\kappa}{4} \nabla^{-4}\nabla_i {\cal P}^i \hspace{1cm} x^i_{iso} = x^i_{ADM}- \frac{1}{4} \nabla^{-4} \partial^i {\cal E}.  \label{isoadm}
\eey
The relation between the coordinates depend on the matter degrees of freedom. In classical Geometrodynamics, the  Hamiltonian descriptions  in the two gauges   are equivalent, because the spacetime metric can be unambiguously interpreted in terms of evolving (Heisenberg-type) canonical observables \cite{ADM}. Both Hamiltonian systems share a common spacetime description.

However, the equivalence of theories at different gauges    is far from guaranteed in quantum theory. First, because there is no  gauge-independent spacetime description that sets a standard of comparison. There is no guarantee that the different Heisenberg evolutions of canonical observables are compatible. In fact, the comparison between different Heisenberg evolutions make no sense. Consider, for example, Eq. (\ref{isoadm}) that connects the coordinates between the two gauges. In quantum theory, energy and momentum densities are operators, while spacetime coordinates are by definition $c$-numbers; there is no mathematical way of connecting the different evolutions. On physical grounds, one could say that the existence of quantum fluctuations in energy and momentum densities makes the relation between reference frames ambiguous.

This problem had been already pointed out by Arnowitt, Deser and Misner \cite{ADM}. More recent work on PFTs confirms that Fock spaces  like the one employed for quantization in the ADM gauge are too "small" to contain unitary transformations between different foliations \cite{ToVa2}. While the analysis of \cite{ToVa2} involves free scalar fields, in which full mathematical control is possible, it is natural to expect its generalization to a broader class of field theories.

These results imply that an analysis of gravitational phenomena in quantum systems that employs Newtonian concepts has no unique spacetime picture. Different gauge choices allow for different embeddings of the Newtonian potential in a spacetime   metric, by using different gauge choices.  Each gauge choice provides an inequivalent way of constructing a spacetime  observables from the quantum description, with no way to choose a `correct' one, if such exists. This ambiguity is a severe problem to all  attempts to provide a spacetime description for gravitational quantum phenomena.

In our opinion, the natural resolution of this problem is to construct a quantum PFT and avoide gauge fixing altogether, as we explain in the next section. However, it is conceivable that a  restricted set of quantum observables could preserve the classical equivalence between Hamiltonian evolution at different gauges, at least in the Newtonian regime. Hence, at least some physical predictions could turn out to be gauge independent.  For this reason, it is important to explicitly construct physically relevant quantum observables, like (\ref{tauobs}) at different gauges. This issue is currently under investigation.


The problem described above is not specific to the weak gravity set-up. It is a facet of a well-known problem in quantum gravity, the {\em problem of time}, and it originates from the incompatible ways time is treated in GR and in quantum theory. It is a structural problem, and, hence, it does not go away when gravity is weak, the same way it does not go away when using other approximations, like minisuperspace models. The interesting thing here is  that the problem is manifested at systems that are experimentally accessible, suggesting that at least some ideas about the solution to the problem of time could be amenable to experimental tests.

 Note that the analysis of gauge fixing in this section proceeds according to the ideas of Arnowitt-Deser-Misner that gauge fixing is equivalent to the choice of a coordinate system, and subsequent analyses that such procedures could provide a solution to the problem of time \cite{Isham92}. This
treatment is manifestly different from the common analysis of gauge fixing in systems with first class constraints that employs ghost fields and techniques like the Becchi-Rouet-Stora-Tyutin (BRST) \cite{HeTe}  or the Batalin-Fradkin-Vilkovisky \cite{FaVi, BaVi} (BFV) formalisms. These techniques
primarily focus on deriving an anomaly-free perturbation theory, and they work at the level of path integrals. In particular, the BFV formalism can be used in order to construct formal  path integrals for the propagator \cite{Teitel, BaPo, Halliwell}. However, in this method   the connection of gauges with coordinate systems, which is the basis of ADM gauge analysis is lost; the resulting propagator carries no memory of the initially chosen gauge choice.
In  fact, application of the BFV method  to GR leads to a Wheeler-DeWitt equation for the quantum state \cite{Halliwell}, and in this sense it is closer to the Dirac quantization of Sec. 5, rather than the reduced state space quantization that is studied in this section.

\section{Dirac quantisation of parameterised field theory}
 We saw that reduced phase space quantisation leads to an effective field theory that reproduces the results of non-relativistic Newtonian theory. The price is that the spacetime description is lost because of gauge fixing. In this section, we describe Dirac quantisation of the system, i.e., the implementation of the constraints after quantisation.

In full quantum gravity, approaches that are based on Dirac quantisation also suffer from the problem of time. However, the linearised theory has a crucial characteristic that is not shared by the full gravity theory.  As shown in Sec. 3, the constraints are linear with respect to the momenta $p_{\tau}$ and $\pi_{\chi}^i$, and the classical   theory is a PFT.

To understand the implications of this point, we note that an influential approach to the problem of time by Kuchar  seeks a canonical transformation that  brings  GR in the form of a PFT. If this were possible,   then Dirac quantisation could lead to a quantum theory that incorporates time evolution with respect to all different spacelike foliations, since those foliations would be defined in terms of geometrodynamic variables \cite{Kuchar92}. This approach fails in full GR, because the desired canonical transformation does not exist,
 even though it can be found (rather artificially) in  presence of some types of matter, like dust \cite{BrKu95}.

However, linearised gravity with matter is already a PFT. If it can be quantised consistently, then it could provide an unambiguous definition of spacetime observables and a consistent description of time evolution with respect to all foliations.  The problem here is that quantum PFTs are difficult to construct, and they may even not exist \cite{Kuchar88, Kuchar89, ToVa1, ToVa2, Var06, LaVa}. Even quantum PFTs that can be constructed are very different from ordinary  QFTs

It is not the aim of this paper to develop this theory in detail. This would require a different publication, as such a construction would involve many technical and conceptual details  beyond the state-of-the-art on the topic. Here, we will only present the key ideas, and we will explain why their rigorous implementation could lead to a theory that is predictive and it provides an unambiguous definition of spacetime observables.

In Dirac quantisation, we first quantise the unconstrained systems, that is, we define a Hilbert space ${\cal H}_0$, where the fundamental field variables exist as operators, subject to the standard commutation relations.
Hence, we have operators $\hat{\cal X}^{\alpha} (x)$, $\hat{w}_{ij}(x)$, $\hat{q}_a(x)$, $\hat{P}_{\alpha}(x)$, $\hat{\pi}_w^{ij}(x)$ and $\hat{p}^a(x)$, such that
\begin{eqnarray}
[\hat{\cal X}^{\alpha}({\pmb x}), \hat{P}_{\beta}({\pmb x}')] = i \delta^{\alpha}_{\beta} \delta^3({\pmb x}, {\pmb x}'),
\eey
\bey
  [\hat{w}_{ij}({\pmb x}), \hat{\pi}_w^{kl}({\pmb x}')] = i (\delta_i^k \delta_j^l + \delta_i^l\delta_j^k) \delta^3({\pmb x}, {\pmb x}'),
\end{eqnarray}
\bey
 [ \hat{q}_a({\pmb x}), \hat{p}^b({\pmb x}')] = i \delta_a^b\delta^3({\pmb x}, {\pmb x}').
\end{eqnarray}
Note that, for simplicity, we switched to the covariant-like notation of Sec 3.3.

Then, on the Hilbert space ${\cal H}_0$, we define self-adjoint operators $\hat{\cal G}^{\alpha}$ that represents the functions ${\cal G}^{\alpha}$ of the canonical variables.
The quantum constraints correspond to the operators $\hat{\cal C}^{\alpha}$, where ${\cal C}^0: = \hat{\pi}_{\tau} +\hat{\cal G}^0$ and  ${\cal C}^i: = \hat{\pi}^i_{\chi} +\hat{\cal G}^i$. The physical Hilbert space ${\cal H}_{phys}$ consists of all states $|\Psi_{phys}\rangle$ that are annihilated by the constraints,
\bey
\hat{\cal C}^{\alpha} |\Psi_{phys}\rangle = 0 , \label{diracq}
\eey
and physical observables  correspond to operators $\hat{A}$  that commute with the constraints $[\hat{A}, {\cal C}^{\alpha}] = 0 $.

There are   complications regarding the definition of the inner product on ${\cal H}_{phys}$, but heuristically, we can work with the Schr\"odinger representation. For simplicity, we focus in the regime where gravitational waves are negligible, so we drop all terms involving $w_{ij}$ and $\pi_w^{ij}$.  Then,
elements of ${\cal H}_0$ are represented by wave functionals $\Psi[{\cal X}^{\alpha}, q_a]$; the variables ${\cal X}^{\alpha}$ and $\hat{q}_a$ act multiplicatively on $\Psi$; the momenta are $\hat{P}_{\alpha}(x) = - i \frac{\delta}{\delta {\cal X}^{\alpha} (x)}$,  and $\hat{p}^a(x) =  - i \frac{\delta}{\delta q_a(x)}$. Then, Eqs. (\ref{diracq}) become
\bey
i \frac{\partial}{\partial {\cal X}^{\alpha}(x)}  \Psi[{\cal X}^{\alpha}, q_a] = {\cal G}^{\alpha}\left({\cal X}^{\alpha}, q_{\alpha}, - i \frac{\delta}{\delta q_a(x)}\right)\Psi[{\cal X}^{\alpha}, q_a]. \label{FSE}
\eey
Hence, the Wheeler-DeWitt equation that is typically obtained by Dirac quantisation becomes a functional Schr\"odinger equation that describes the evolution of the wave functional $\Psi$ along any spacelike foliation. It is remarkable that this happens automatically in weak gravity, we do not need to select the variables that define the foliation, they are straightforwardly obtained by the linearisation procedure.

Eq. (\ref{FSE}) is formal. To obtain physical predictions from it, we must identify a suitable inner product on ${\cal H}_0$ so that the functional Schr\"odinger equation corresponds to a unitary evolution. An inner product will allow us to define probabilities, and thus make concrete physical predictions for weak gravity experiments. It is important to remark that the specific choice of foliation coordinates ${\cal X}^{\alpha}$ made here allows for a consistent spacetime picture, because, they satisfy the consistency condition \cite{Isham92}
\bey
\{{\cal X}^{\alpha}, {\cal H}[L]\} = 0 \; \; \mbox{when} \; \; L(x) = 0, \label{condition}
\eey
where ${\cal H}[L]$ is the Hamiltonian constraint smeared with the test function $L$. By the analysis of \cite{Isham92}, a mathematically rigorous implementation of Eq. (\ref{FSE}) will most likely lead to the definition of consistent spacetime observables.

In absence of gravitational waves, the only degrees of freedom that appear in the wave-functional are matter degrees of freedom and the coordinate parameters ${\cal X}^{\alpha}$. For predictions at weak gravity, it suffices to construct an effective field theory, which will implement the self-gravity contribution perturbatively.

Hence, the natural procedure is to split ${\cal G}^{\alpha} = {\cal G}^{\alpha}_0 + {\cal G}^{\alpha}_{gsi}$, where  ${\cal G}^{\alpha}_0$ corresponds to matter without gravitational self-interaction and ${\cal G}^{\alpha}_{gsi}$ corresponds to the self-interaction terms. The idea is to find a rigorous description for the functional Schr\"odinger equation with ${\cal G}^{\alpha}_0 $,
\bey
i \frac{\partial}{\partial {\cal X}^{\alpha}(x)}  \Psi[{\cal X}^{\alpha}, q_a] = {\cal G}_0^{\alpha}\left({\cal X}^{\alpha}, q_{\alpha}, - i \frac{\delta}{\delta q_a(x)}\right)\Psi[{\cal X}^{\alpha}, q_a], \label{FSE1}
\eey
and then to incorporate ${\cal G}^{\alpha}_{gsi}$ as a perturbative correction.

But then, Eq. (\ref{FSE1}) should coincide with the PFT description of ordinary matter in absence of any gravitational effects. For matter consisting of a free scalar field, this theory has been fully analysed by Torre and Varadarajan \cite{ToVa2}. They showed that there is no unitary implementation of functional evolution in the usual Fock space of the free scalar field (this is possible only in two dimensions). This implies that any quantum theory constructed from Dirac quantisation is unitarily inequivalent to the quantum theory obtained in the ADM gauge, and, hence, it is likely to lead to different physical predictions. This is not unexpected, because, as explained, the ADM theory gives ambiguous results for spacetime observables, while the quantised PFT---if it exists---will provide a consistent spacetime picture.

However, the rigorous construction of a quantum PFT will lead to theories that are very different from ordinary QFTs, where by ordinary, we denote QFTs that are constructed through perturbation theory of Fock-space QFTs. One possibility is to implement Eq. (\ref{FSE1}) using polymer quantisation, i.e., using the techniques developed in relation to the Loop Quantum Gravity (LQG) program. This was achieved by Laddha and Varadarajan in Ref. \cite{LaVa}---see, also Ref. \cite{TT}. The problem with this method is that these representations are not continuous, i.e., one cannot derive the associated generators, including Hamiltonians for translations with respect to specific foliations. Still, polymer quantization is a plausible candidate for a predictive spacetime description of weak gravity that is contiguous to the LQG program. Alternatively, a PFTs could be constructed from path-integrals that are defined with respect to different measures on the space of paths than those that are employed in usual QFTs.

We believe that the most promising approach is {\em histories canonical quantisation} \cite{Sav09}, i.e., a generalisation of canonical quantisation that employs histories as fundamental notions, rather than states defined at a single moment of time \cite{IL, Sav99, Sav04a, Sav04b}. In this framework, a Fock-like quantisation of free fields that could allow for general functional evolution is possible \cite{Sav02, SavIsh}, albeit in a larger Hilbert space whose vectors correspond to histories rather than single-time states \cite{Ish94}. This approach is currently under investigation by one of us (K.S.).

\section{Discussion}

\subsection{Degrees of freedom in the Newtonian regime}
Our analysis has made it obvious that the only degrees of freedom present in the Newtonian regime correspond to matter.   The only gravitational degrees of freedom in weak  gravity pertain to gravitational waves / gravitons. These are {\em decoupled} at the Newtonian limit. Hence, there is no possibility for the experiments proposed in Refs. \cite{Bose17, Vedral17}, or for any other experiments in this regime to provide an experimental proof that gravity is fundamentally quantum.

 This conclusion does not depend on the choice of gauge or on the quantisation method. The identification of true degrees of freedom in a constrained system is an elementary procedure that is not affected by gauge choice or quantisation method. Even in path integral quantisation, the integration measure is essentially defined for paths over paths on the reduced state space.
Gravity in the Newtonian regime involves only a non-local interaction Hamiltonian for matter, and as shown in this paper, the gauge variables $\chi^i$ and $\tau$ that correspond to the independence of the physical predictions on the choice of foliation.

This does not mean that quantum experiments in this regime cannot provide information about the fundamental nature of gravity, indeed, in Sec. 6.3 we do argue that, {\em assuming that gravity is quantum}, experiments in the Newtonian regime may, in principle, allow for a distinction between different quantum gravity theories.

In general, what exactly can be proven experimentally depends on prior assumptions about alternative theoretical explanations. In the present case, we have a  clear idea of what a quantum theory of gravity entails, at least in the Newtonian regime. But what is the alternative theory of classical gravity interacting with quantum matter against which experiments are supposed to test? One popular candidate is the Moller-Rosenfeld theory \cite{Moller, Rosenfeld}, according to which the source in Einstein equations in the expectation value of the stress-energy tensor\footnote{The semi-classical Einstein equation $G_{\mu \nu} = \frac{\kappa}{2} \langle\hat{T}_{\mu \nu}\rangle$  is fundamental in this theory, rather than an approximation to a quantum gravity theory. It is far from obvious that this theory is consistent, or even that it makes mathematical sense, but formal manipulations can lead to predictive evolution equations in the Newtonian regime \cite{NS}.}. In this theory, the full metric $g_{\mu \nu}$ is fundamentally classical; the variable $\phi$ of Eq. (\ref{split}) is not slaved to the quantum matter distribution. It acts like a partially classical channel of interaction that removes information from the quantum state, and hence, it disallows gravity-induced entanglement. Quantum experiments in the Newtonian regime can certainly rule out the Moller-Rosenfeld theory.

Consider, in contrast, the model of Ref. \cite{AnHu13} for the interaction of a quantum system with gravitational perturbations. In this model, the transverse-traceless degrees of freedom are treated as classical stochastic \footnote{The model of Ref. \cite{AnHu13} is  agnostic on whether gravity is fundamentally quantum or not. The stochastic perturbations may originate from the coarse-graining of fundamentally quantum gravitational degrees of freedom.}. However, the constraints (\ref{const1a}) and (\ref{const2a}) still apply. The variable $\phi$ is fully slaved to the quantum matter distribution via the constraints and matter degrees of freedom are described by the quantum Hamiltonian  (\ref{hgsiq}). Decoherence due to gravitational perturbations occurs in the momentum basis, and it
does not, in general, destroy entanglement between static bodies.
Quantum experiments like the ones of Refs. \cite{Bose17, Vedral17} cannot distinguish between this model and a theory that treats  gravitational perturbations quantum mechanically.

It is important to emphasize that
mathematically consistent couplings of quantum to classical variables typically induce non-unitary evolution and decoherence to the quantum system and noise to the classical system \cite{BlJa, HaDi, Diosi3, HaRe05}.
Models for quantum-to-classical coupling without decoherence and noise typically lead to the violation of positivity, i.e., negative probabilities.
This implies that the gravitational field can be consistently coupled to quantum matter only if it is subject to stochastic fluctuations---for models see, \cite{HaRe05, Oppenheim}. To distinguish between classical-stochastic versus quantum fluctuations, we must access  the degrees of freedom that fluctuate. We do not have access to such degrees of freedom for gravity in the Newtonian regime.

In Ref. \cite{claims2}, it is claimed that, when a mass is put into a spatial superposition and interacts
gravitationally with a test mass, one runs either into faster-than-light signalling or violation of quantum complementarity, unless one takes into account vacuum fluctuations of the gravitational field, and the emission of gravitational radiation in a coherent superposition. However, a close reading of the argument shows that vacuum fluctuations need not be quantum, and that the restoration of quantum complementarity only requires a decoherence mechanism---spontaneous emission of discrete quanta being only one of possible scenarios. Hence, the arguments of Ref. \cite{claims2} do not rule out theories in which gravity is treated as a classical stochastic field that causes decoherence to quantum matter, which are properties that any mathematically consistent quantum-to-classical coupling must have, anyway.
Ruling out inconsistent  theories of quantum-to-classical coupling, as in  \cite{claims2},   is of high importance for mapping out the set of theories that can be experimentally tested.
However,  it does not provide evidence that gravity is quantum, unless all non-quantum theories of gravity are proven inconsistent.

Our analysis shows that, {\em according to GR}, the two parts of a quantum bipartite system that interact gravitationally in the Newtonian regime {\em do so without an intermediate degree of freedom}. In general, this interaction leads to the creation of entanglement and other manifestations of quantum behavior.
Obviously, if  an intermediate degree of freedom existed, the measurement of gravity-induced entanglement would show that this degree of freedom were quantum. However, the very postulate of such a degree of freedom contradicts GR.

\subsection{Locality and observables}
The classical Hamiltonian that describes the gravitational interaction in the Newtonian regime is non-local in all gauges with respect to the matter fields. This is not a pathological feature and certainly not an indication of action-at-a-distance. It is also not an artifact of the non-relativistic approximation, as such terms appear in the full linearised theory---see, Eq. (\ref{hadm}).
The non-locality of the Hamiltonian is a prediction of GR, and   it is fully compatible with the causality and locality that characterises classical GR.

The reason for this behavior of the Hamiltonian is the fact that GR is subject to first-class constraints. In such systems, the Hamiltonian is uniquely defined only at the constraint surface. However, to restrict to the constraint surface one has to solve Eq. (\ref{constraint1}) for the potential $\phi$, and this leads to the appearance of non-local terms in the Hamiltonian. Exactly the same happens in the Hamiltonian treatment of the electromagnetic field \cite{WeinbergQFT}, as well as in its subsequent quantisation. In classical physics,  constraints are {\em instantaneous laws} and not equations of motion; they restrict the initial data at each equal time surface, they do not evolve these data.

The existence of the constraints implies non-local correlations between the true degrees of freedom\footnote{
In general, the true degrees of freedom are non-local functionals of the Lagrangian (covariant) fields. This fact is often obscured in path integral treatments through the use of local gauges, but  the non-locality is expressed the Gribov ambiguity \cite{Gri}. There are  manifestations of this non-locality even in linear systems \cite{Strocchi}. }. We believe that this requires an update of usual notions of local operations when quantum systems interact gravitationally: gravity cannot be switched off and it always leads to correlations between observables at separated regions.

Consider, for example,  gravity-induced entanglement by two masses $m_1$ and $m_2$. Let us assume that both masses $m_i$ are prepared within the gravitational field of the Earth. The interaction with the Earth does not affect the dynamics of any experiment, as long as it does not involve significant motion in the vertical direction. Nonetheless, if the quantum system has an extension $d$ in the vertical dimension, the gravitational interaction of a mass with Earth---given by Eq. (\ref{hint})---is of order $mgd$, where $g$ is the gravitational acceleration of the Earth. By perturbation theory, this leads to an entanglement of order $mgd/E$ between the mass and Earth's degrees of freedom. When tracing out the degrees of freedom of the Earth, the reduced density matrix for the small mass will be mixed with an entropy of the order
 $S = mgd/E$. For microscopic systems, this degree of mixing is negligible, e.g., for the hydrogen atom $S \sim 10^{-17}$.

However, this mixing is stronger in macroscopic quantum systems. When the two masses are brought at distance $L$, the induced entanglement is of the order of $\alpha = \frac{Gm_1m_2}{LE}$. For $m \sim 10^{12}amu$, and $d, L \sim 100 nm$, $S/\alpha \sim 10^{13}$.   It seems unlikely that the interaction between the two masses can generate significant entanglement, all effects would be drowned by the mixed-ness of the initial state\footnote{Note that this result is independent of the energy $E$ of the macroscopic quantum system. It follows only from the fact that the interaction term of each subsystem with Earth  is much larger than the Newtonian attraction of the two subsystems.}. This is equivalent to saying that the matter degrees of freedom of the Earth can act as sources of environment induced decoherence through the channel of the Newtonian force. There may exist ways to shield against such decoherence, i.e.,  to find decoherence-free subspaces,  or to carry out such experiments in space.

Nonetheless, this example demonstrates the pitfalls in ignoring the persistent correlations generated by gravitational fields even if they do not affect the dynamics of the system.
  In our opinion, dynamical effects that go beyond correlations---like Rabi oscillations---are more robust, and they are better candidates for experimental demonstrations of gravitational effects in macroscopic quantum systems.

\subsection{A spacetime covariant theory for weak gravity}
In Sec. 4, we showed that quantisation in the ADM gauge leads to an effective QFT description that leads to the expected results in the Newtonian regime. The price one has to pay is that the spacetime picture is lost: there is no guarantee that spacetime observables are gauge independent; in fact, most probably they are not.  We argued that there is a natural resolution to this problem, namely, the quantisation of gravity as a PFT. The problem here is that quantum PFTs in dimensions more than two do not have any Fock-type representations, hence, if they exist, their structure is very different from that of weak gravity in the ADM gauge.

 This likely implies that the quantum PFTs will lead to different predictions for at least some observables. It is a natural conjecture that the difference will be most pronounced in spacetime observables, like path proper times or observables that implement the ordering of events.

 The construction of  a predictive and spacetime covariant quantum description is a challenge for any program to quantum gravity. First, the derivation of a  theory through which quantum experiments in weak gravity are described---including predictions about spacetime observables---is a non-trivial check for all research programs. Second, it is likely
  that different approaches to quantum gravity will lead to different predictions for macroscopic quantum phenomena, at least as far as spacetime observables are concerned. Such predictions might not be testable with existing experimental capabilities. However, the construction of controlled quantum systems at increasingly larger macroscopic scales appears much easier than accessing Planck-scale energies; deep space quantum experiments are particularly promising \cite{dsql}. Third, we showed that structural problems of quantum gravity, like the problem of time and the problem of observables, persist at the weak gravity regime. Any proposed solution should not only work for Planck scale physics, or in quantum cosmological models, but also for weak gravity. In the medium term, experiments in weak gravity could help winnow the field of candidate solutions to those problems.

  In our opinion, the usual methods of canonical quantisation will not work with the quantisation of PFTs. We need a quantisation procedure that incorporates both the insights of the canonical approach about  observables and  constraints and a genuine spacetime description. This is possible in the histories description of canonical gravity \cite{Sav04a, Sav04b}, and in the associated quantisation scheme that is based on histories \cite{Sav09, SavAn00, SavAn06}.

\subsection{Decoherence of the Diosi-Penrose type}
Penrose's proposal of gravitational decoherence is essentially motivated by the problem of time in quantum gravity \cite{Penrose}. The contradiction between the role of time in GR and in quantum theory is manifested clearly when considering superposition of macroscopically distinct states. Each component of the superposition generates a different spacetime. Since there is no canonical way of relating time parameters in different spacetime manifolds, there is a fundamental ambiguity in  the time parameter of the evolved quantum state. According to Penrose, this ambiguity is manifested even at low energies, when the gravitational interaction can be effectively described by the Newtonian theory. It leads to  mechanism of gravity-induced decoherence for superpositions of  states with different mass densities $\mu_1(x)$ and $\mu_2(x)$. Penrose proposed that the time-scale of decoherence is of the order of   the gravitational self-energy of the difference in the two mass densities.

Penrose's rationale for gravitational decoherence  likely fails if one can construct a quantised PFT for weak gravity, as described in Sec. 5. In this theory, it would be possible to describe time evolution in terms of the variable $\tau$ that is intrinsic to the  metric, and not an arbitrary coordinate of the spacetime manifold. In contrast, if such a construction turns out to not be possible, then Penrose' s argument receives additional support.

Even in this case, we think that the identification of the gravitational self-energy as the order of magnitude for decoherence is not justified by our analysis of the weak gravity limit. As shown in Secs. 2 and 3, the gravitational self-energy term is robust, i.e., it appears the same in all foliations. By Eq. (\ref{G0N}), the foliation-dependent contribution to the energy arises from a total divergence.
 There is no preferred value for this quantity in our analysis. The natural decoherence time scale would most probably arise out of unknown physics, and there is no physical justification why it should be related to gravitational self-energy. As shown in slightly different contexts \cite{AnHu08, AnHu13}, GR does not contain a natural time scale that can be interpreted as deoherence time.

\subsection{Gravitational waves and the noise of gravitons}
In this paper, we focused on the Newtonian  regime and ignored the gravitational wave perturbations. Still, the formalism presented here also applies  to quantum systems interacting with gravitational waves.

There are two classes of experiments involving gravitational waves where our analysis is relevant. First, in the use of macroscopic quantum systems as gravitational-wave detectors---see, for example, \cite{MMM}. This case does not require an analysis of the backreaction of the quantum system to the wave. The gravitational wave can be treated as a classical external force as is common in gravitational wave detection. As long as we refrain from discussing spacetime observables, this set-up can be fully analysed in the ADM gauge.

On the other hand, there is the possibility that the incoming wave contains quantum contributions, for example, it may correspond to amplified quantum fluctuations during inflation. It has been proposed in Ref. \cite{WZ} that the noise induced by such waves on gravitational wave detectors would separate between classical and quantum waves. Existing models focus on a path-integral analysis, using a fixed gauge for gravitons \cite{WZ2, Haba, KST},  and then employ a semi-classical approximation to identify the noise. There are  potential pitfalls in such analyses, for example, the problems with gauge dependence discussed in Sec. 4,  or the consistent definition of noise.

We think that experiments measuring  quantum effects from gravitons are plausible.  Such experiments would truly test the quantum nature of gravity, as they are not restricted in the Newtonian regime, but they access true degrees of freedom. However, concrete proposals for such experiments require    sharp criteria for the distinction between quantum and classical that can only be obtained from a first-principles
analysis similar to the one proposed here.

\section*{Acknowledgements}
C. A. would like to acknowledge many discussions with B. L. Hu on the topic. This research is co-financed by Greece and the European Union (European Social Fund- ESF) through the Operational Programme "Human Resources Development, Education and Lifelong Learning 2014-2020" in the context of the project “Influence of Gravity on Quantum Entanglement” (MIS 5047122).”

\begin{appendix}

\section{Parameterised field theory}
Consider a field theory on Minkowski spacetime described by the variable $q(x, t)$ and its conjugate momentum $p(x, t)$. For simplicity, we drop any indices on $q$ and $p$. The first-order  action is
\begin{eqnarray}
S = \int dt d^3x (p \frac{\partial q}{\partial t} - {\cal  H}(q, p)).
\end{eqnarray}
We undertake a spacetime parameterisation of the action. That is we promote the coordinates $x$ and $t$ to variables that depend on other spacetime coordinates $\tau$ and $y$, and we define the fields with respect to the latter. To this end, we write $t(\tau, y)$ and $x(\tau, y)$. Then, $dtd^3x = \dot{t} L d\tau d^3y$, where a dot denotes derivative with respect to $\tau$ and $L:=  \det (\partial x^i/\partial y^j)$. Then,
\bey
S = \int d\tau d^3y (L p \dot{q} - L p \frac{\partial q}{\partial x^i}\dot{x}^i - L\dot{t} {\cal H}(p, q, y, \tau)).
\eey
We define $\bar{p} = Lp$, and
\bey
\pi^0 := -L {\cal H}(p, q,y, \tau)\hspace{1cm}
 \pi_i := -Lp \frac{\partial q}{\partial x^i}. \label{constrp}
\eey
Then, $S = \int d\tau d^3y (\bar{p} \dot{q} + \pi_i \dot{x}^i+ \pi_t \dot{t}$, subject to the constraints (\ref{constrp}). The latter are implemented by incorporating them into the action through Lagrange multipliers $N^0, N_i$,
\begin{eqnarray}
S = \int d\tau d^3y \left[\bar{p} \dot{q} + \pi_i \dot{x}^i+ \pi_t \dot{t} - N^0(\pi_t +L {\cal H}) - N^i(\pi_i + Lp \frac{\partial q}{\partial x^i}) \right]. \label{actpar}
\end{eqnarray}
The variation of Eq. (\ref{actpar}) with respect to the Lagrange multipliers yields Eq. (\ref{constrp}).

The symplectic form associated to $S$ is
$\Theta:= \int d^3y \left(\bar{p}\delta q + \pi_i \delta x^i + \pi_t \delta t\right)$. On the constraint surface,
\bey
\Theta:= \int d^3y \left( \bar{p}\delta q -Lp \frac{\partial q}{\partial x_i} \delta x^i -L {\cal H} \delta t \right). \label{parameterized}
\eey

\end{appendix}

\end{document}